# Rotation of Comet Hartley 2 from Structures in the Coma

Nalin H. Samarasinha[1,3], Beatrice E.A. Mueller[1,3], Michael F. A'Hearn[2], Tony L. Farnham[2], and Alan Gersch[2,3]

[1] Planetary Science Institute, 1700 E Fort Lowell Road, Suite 106, Tucson, AZ 85719, USA; nalin@psi.edu

[2] Department of Astronomy, University of Maryland, College Park, MD 20742, USA

---

[3] Visiting Astronomer, Kitt Peak National Observatory, National Optical Astronomy Observatory, which is operated by the Association of Universities for Research in Astronomy (AURA) under cooperative agreement with the National Science Foundation.

# ABSTRACT

The CN coma structure of the EPOXI mission target, comet 103P/Hartley 2, was observed during twenty nights from September to December 2010. These CN images probe the rotational state of the comet's nucleus and provide a ground-based observational context to complement the EPOXI observations. A dynamically excited cometary nucleus with a changing rotational rate is observed, a characteristic not seen in any comet in the past. The lack of rotational damping during the four-month observing interval places constraints on the interior structure of the nucleus.

# 1. INTRODUCTION

The rotational state of a comet's nucleus is a basic physical parameter needed to accurately interpret other observations of the nucleus and coma. Furthermore, changes in the rotational state provide strong constraints on the nuclear activity and serve as a probe of the interior structure of the body. Because the rotational state of the nucleus is difficult to measure directly, indirect methods must be used, and coma morphology (e.g., a spiral produced by a jet of material flowing outward from a spinning body) has proved to be a very useful technique (Farnham 2009). In addition to reflecting the rotational state of the nucleus, the morphology and evolution of features can be used to understand the comet's activity and the physics and chemistry of the coma. The temporally varying spatial features in CN, first detected in comet 1P/Halley (A'Hearn et al. 1986), have been particularly useful for this task. The CN spatial features imaged using a CN filter monitor the flux due to fluorescence emission from the CN violet (0-0) band.

The CN features in the coma of comet 103P/Hartley 2 show clear temporal and spatial variations over timescales as small as a few hours as well as over much longer timescales, for example, between observing runs (Figure 1). Analysis of these features indicates that the nucleus is spinning down, and suggests that it is in a state of a dynamically excited rotation. Although CN morphology has been used to constrain the rotation of a number of comets (e.g. 1P/Halley (Samarasinha et al. 1986; Hoban et al. 1988), Machholz (C/2004 Q2) (Farnham et al. 2007)), it has never been previously used to infer a changing rotational state.

In this Letter, we present observational evidence that the CN coma morphology of 103P/Hartley 2 indicates an excited rotational state of the nucleus that varies with time. The results from a numerical integration is shown to demonstrate that a highly active, small, and elongated nucleus like 103P/Hartley 2 could indeed exhibit such a rotational behavior. Finally, lack of rotational damping is used to place constraints on the interior structure of the nucleus.

# 2. OBSERVATIONS

We obtained images of 103P/Hartley 2 at the 2.1-m telescope at the Kitt Peak National Observatory from September 1-3, September 30-October 4, November 2-8, and December 11-15, 2010 UT. The corresponding heliocentric distances to the comet during each of these observing runs were 1.30, 1.12, 1.06, and 1.23 AU, while the geocentric distances were 0.38, 0.18, 0.16, and 0.35 AU. The respective solar phase angles (hereafter α) were 34°, 45°, 59°, and 39°. The sun and Earth directions as seen from the comet have changed 28° and 26° respectively between the first and the second observing runs. The corresponding changes between the second and third runs were 39° and 88° whereas the changes between the third and fourth runs were 40° and 21°. This means that while the changes in the sun direction between consecutive observing runs are comparable, the change in the Earth direction between the second and third run is much more extreme than those between the first and second runs or between the third and fourth runs.

Our observing runs bracketed the comet's perigee (0.12 AU) on October 20, its perihelion passage on October 28, and the EPOXI encounter on November 4. The comet was observed with a blue-sensitive CCD and the field-of-view is about 6.5×10 arcmin in the north-south and east-west directions respectively. The pixel scale is 0.305 arcsec pixel$^{-1}$ and corresponds to 84, 39, 35, and 78 km pixel$^{-1}$ for the September, October, November, and December runs respectively. The images were taken with broadband R as well as narrowband comet filters (Farnham et al. 2000). This Letter focuses on the narrowband images that isolate gaseous CN, with which we were able to achieve high signal-to-noise. The CN filter has a central wavelength of ~3869 Å and a full-width-at-half-maximum of ~56 Å (Farnham et al. 2000).

## 3. DATA ANALYSIS AND RESULTS

All the CN images were bias subtracted and flat-fielded. Each image was then divided by its own azimuthally averaged profile to enhance the CN features present in the coma. This technique (Schleicher & Farnham 2004), effectively removes the gross background, thereby enhancing the features above the background coma. Due to the low dust-to-CN ratio for this comet (Schleicher 2010), the dust contamination in the CN filter is negligible and has no effect on the CN coma morphology. This was confirmed by comparing the features in the continuum images (which are dominated by scattering from the dust and icy grains), taken immediately prior to or after the CN images, with the features in the CN images. Finally, the absence of a feature in the CN images that is co-spatial with the dust tail provides additional evidence that the dust contamination is not an issue.

Figure 2 shows the CN morphology during our September run. The morphology indicates a clockwise rotation of the nucleus that forms an Archimedean spiral-like pattern in the coma. The temporal evolution of this jet-like feature from the north to the west, then to the south of the nucleus, and finally to the east is clear in the images. The apparent outflow velocity is larger in the north than in the south, indicating that the jet is closer to the line-of-sight when it is in the south. Furthermore, the clockwise movement of the feature indicates that the rotational angular momentum vector[4] (hereafter RAMV) is >90° away from the Earth direction. The best repeatability of the features is seen for a periodicity near 17.1 hours (Samarasinha et al. 2010).

---

[4] Strictly speaking, we can only make inferences about the direction of the spin vector; however, as the morphology is consistent with a nucleus that effectively shows little deviation from a principal-axis rotation around the axis of maximum moment of inertia, the inferences are also valid for the RAMV direction.

The morphology has not radically evolved between the September and October runs, but distinct differences were observed (not shown). For example, the spiral feature from September now barely moves to the south of the line-of-sight and it is still produced by clockwise rotation. Evidence of a second, weaker feature, partially overlapping with the main one, is seen in the east/northeast (subtle indications for a weak second feature were also seen during the September run). The spiral feature suggests that its source is located on the negative hemisphere of the nucleus (i.e., negative with respect to RAMV) and the RAMV is still >90° away from the Earth direction but by a smaller amount than in September. The morphology and its evolution from the September and October runs are compatible with the preliminary estimate for the direction of the RAMV of (RA=345°, Dec=-15°) with an uncertainty of ~20° (Samarasinha et al. 2010). This estimate was derived based on the position angle which nearly bisects the spiral jet feature (Samarasinha et al. 2010) and the above constraints for the direction of the RAMV described for the September and October runs. Currently complementary data sets by others are not publicly available and comparisons have not yet been carried out.

The best repeatability of the features for the October run is for a periodicity of ~17.6 hours, which is clearly incompatible with the 17.1-hour periodicity derived for the September run (Figure 3). However, as shown in Figure 4, the morphology has small but notable variations in the features between images of the same rotational phase on some consecutive cycles, which suggests a slightly excited non-principal axis rotation. A non-principal axis rotation (colloquially referred to as tumbling motion by some) is a dynamically excited state of the nucleus where rotation is characterized by two independent periods and three component motions (Landau & Lifshitz 2003; Samarasinha & A'Hearn 1991). Bodies with principal-axis rotation exhibit near-perfect repeatability of coma features as seen for example in Machholz (C/2004 Q2) (Figure 5 in Farnham et al. 2007).

By the November run, the CN morphology has changed significantly, at least partly due to the 88° change in the Earth direction since the October run. During the seven nights in November, two almost spatially opposite features (e.g., the November image in Figure 1), one to the north/northeast and one to the south, characterize the CN coma structure and no Archimedean spiral-like features are seen. This morphology is hard to reconcile with the September/October estimate of the RAMV, suggesting that this vector may be migrating due to forced precession induced by reactive torques from the nucleus' outgassing. At this time, the coma morphology repeats with a periodicity ~18.8 hours. However, there are again recognizable differences between certain image pairs of the same rotational phase from different cycles, indicating an excited rotational state.

During the December run, the morphology shows two features — one primarily in the northeast with little or no curvature and another that starts out to the east, moves to the south and then west (e.g., the December image in Figure 1) indicating counter-clockwise rotation of the nucleus. There appear to be changes to the morphologies from night to night, but because the comet was observable for only about five hours per night, it is not clear whether these changes are the result of a non-principal axis rotation or if the southern feature is produced by different active regions on different nights or both. However, the persistent counter-clockwise nature of the southern feature suggests that the RAMV is now <90° from the Earth direction.

For all of the observing runs, the synodic effects are < 0.1 hours. Therefore, the synodic effects cannot explain the above changes for the periodicity of coma repeatability.

## 4. DISCUSSION

Our observations of the CN coma morphology and its repeatability clearly show that 103P/Hartley 2's effective rotation period[5] (cf. Figure 5(a)) has increased during the observing window. Furthermore, there are indications that the nucleus is in an evolving non-principal axis rotational state that includes a likely forced precession of the RAMV. Our results are also consistent with the orientation of the long axis of the nucleus as observed during the EPOXI encounter (A'Hearn et al. 2011). Using order of magnitude calculations and numerical modeling we demonstrate that the above scenario is indeed possible for 103P/Hartley 2.

The timescale, $\tau_P$, for changing the rotation period by torques generated by outgassing is expressed by $\tau_P = \omega/|d\omega/dt| = I\omega/(frVQ_m)$ where $\omega$ is the angular velocity, $r$ is the effective radius of the nucleus, $I$ is the moment of inertia, $V$ is the gas outflow velocity, and $Q_m$ is the mean rate of direct gas sublimation from the nucleus (Samarasinha et al. 1986). The effective moment arm, $f$, required to induce changes in the rotation period is dimensionless and $1 > f > 0$ but the actual value is likely to be near 0 due to cancellation of torques (Jewitt 1997). By adopting $I \sim 2 \times 10^{16}$ kg m$^2$, $\omega \sim 10^{-4}$ s$^{-1}$, $f \sim 0.01$, $r \sim 5 \times 10^2$ m, $V \sim 5 \times 10^2$ m s$^{-1}$, and $Q_m \sim 5 \times 10^1$ kg s$^{-1}$ with $Q_m$ representing only the direct sublimation from the nucleus, we obtain $\tau_P \sim 6$ months — a timescale consistent with the CN morphology.

Note that the timescale for changes in the direction of the RAMV is also of the same order (Samarasinha et al. 1986; Whipple & Sekanina 1979). Therefore, over a timescale of a few months, especially when the activity is high near perihelion, the direction of the RAMV could undergo changes as large as a few tens of degrees.

Figure 5(b-d) illustrates the results of a numerical simulation where a 103P/Hartley 2-like nucleus undergoes rotational excitation accompanied by changes in the direction of the

---

[5] The "effective rotation period" here refers to the periodicity of repeatability of coma features and is identified with the period of the circulatory motion of the long axis (axis of the minimum moment of inertia) around the RAMV.

RAMV. If the net torque is dominated by activity at the ends of the long axis, then the component of the net torque parallel to the long axis is smaller than the components parallel to the short and intermediate axes and the nucleus is unlikely to be highly excited (Gutierrez et al. 2003) — i.e., Long Axis Modes that are highly excited are unlikely. If the activity dominates at other locations, the nucleus could be in such an excited state but the morphology would not repeat in the manner observed in 103P/Hartley 2 nor would one be able to self-consistently explain the radar observations (e.g., Harmon et al. 2010).

The nucleus is in a dynamically excited rotational state that changes with time suggests that the damping timescale (Burns & Safronov 1973), $\tau_{damp}$, for the excited rotation is >6 months. I.e., $\tau_{damp}=K_1\mu Q/(\rho r^2 \omega^3)$ >6 months where $K_1$ is a dimensionless scaling coefficient (Sharma et al. 2005) of the order of 100, $\mu$ is the rigidity (shear modulus), Q is the quality factor (a dimensionless measure of the inefficiency of energy loss due to mechanical deformations.), and $\rho$ is the bulk density of the nucleus. By assuming $\rho \sim 5\times 10^2$ kg m$^{-3}$, r$\sim 5\times 10^2$ m, and $\omega \sim 10^{-4}$ s we obtain $\mu Q$ >10 kg m$^{-1}$ s$^{-2}$. This lower limit is many orders of magnitudes smaller than what has been derived for asteroids (Margot et al. 2002). If the existing early pre-perihelion light curve observations of the nucleus (Meech et al. 2009) or future pre/low-activity pre-perihelion observations of 103P/Hartley 2 could confirm a relaxed principal-axis rotation, then the excited rotational state should damp out over the course of the comet's orbital period. This would provide additional constraints, so that $10^2$ kg m$^{-1}$ s$^{-2}$ > $\mu Q$ > 10 kg m$^{-1}$ s$^{-2}$. If this is the case, then 103P/Hartley 2 would be a flexible and elastic object widely different from a rigid body.

The increase of the effective rotation period during the current apparition suggests that in the past, it could have been sufficiently small to cause a splitting event of the nucleus. This assumes that the component torques due to outgassing during the past apparitions were similar to those of the current apparition. However, this may not necessarily be accurate considering the impact of the high level of activity on the shape of this small nucleus as well as the likely temporal changes in activity over orbital timescales.

## 5. CONCLUSIONS

We show that
• Comet 103P/Hartley 2 is in a low-excitation non-principal axis rotational state that is changing with time
• The effective rotation period is increasing and the changes are easily detectable in a timescale of one month
• The non-gravitational torques are large enough to change the direction of the RAMV by a few tens of degrees during a few months
• The damping timescale for the exited rotation is >6 months, and
• The quantity $\mu Q$ is >10 kg m$^{-1}$ s$^{-2}$ where $\mu$ is the shear modulus and Q is the quality factor.

B.E.A.M. and N.H.S. acknowledge support from the NASA Planetary Astronomy Program. M.F.A., T.L.F., and A.G. were supported partly by UMd and partly by the EPOXI project, NASA contract NNM07AA99C to the UMd. We thank the referee for the comments, which improved the paper. This is PSI Contribution Number 505.

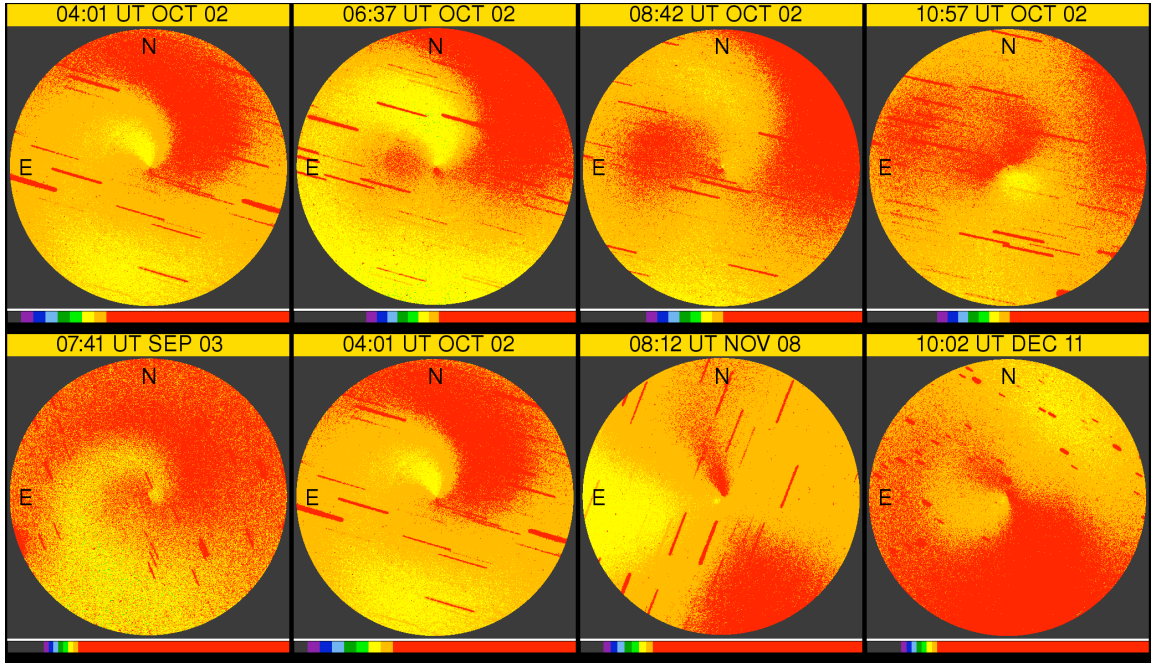

**Figure 1.** CN images showing variations of morphology in a single night (top row) and in different runs (bottom row). Each image was divided by its azimuthally averaged profile to enhance the CN features. The red color denotes brighter regions. The nucleus is at the center of each image. North is up and east is to the left. Each panel is 1,000 pixels across resulting in linear dimensions of ~84,000, ~39,000, ~35,000, and ~78,000 km respectively for the September 3, October 2, November 8, and December 11 images. Times when the images were taken are shown at the top of each image. The streaks are star trails. The sky-plane position angles of the sun measured from north through east (hereafter PA) for September 3, October 2, November 8, and December 11 are 2°, 7°, 108°, and 138° respectively. The respective solar phase angles, $\alpha$, are 35°, 45°, 59°, and 40°.

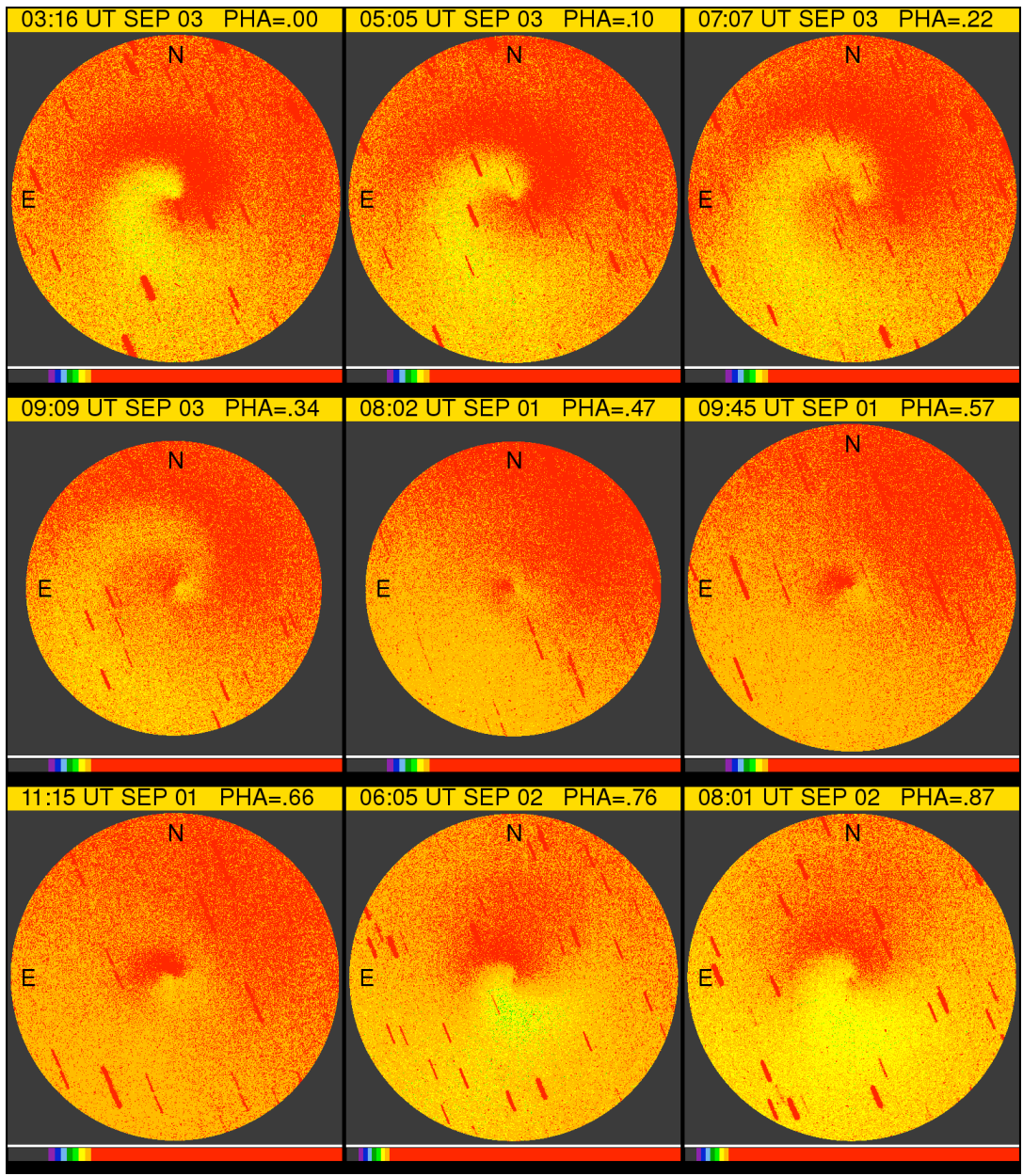

**Figure 2.** CN images from left to right and top to bottom show a rotational sequence from the September observing run phased to a period of 17.1 hours (with the rotational phase listed at the top right of each panel; zero phase was arbitrarily chosen as 00:00 UT on September 1). The images are 1,000 pixels across except the first two images of the middle row, which are 900 pixels across (i.e., ~84,000 versus ~76,000 km across). Note the gradual progression of the spiral feature as a function of the rotational phase. Clouds affected the signal-to-noise of images on September 1. The PA and α are ~3° and ~34° respectively.

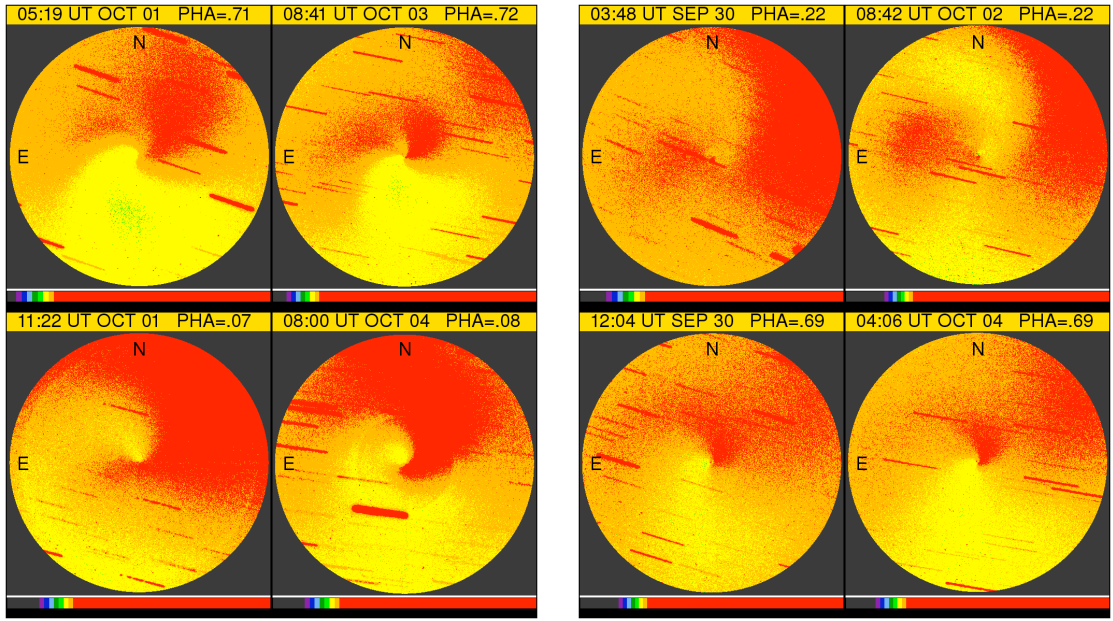

**Figure 3.** Left: two CN image pairs to demonstrate that a periodicity of 17.1 hours is inconsistent with the October run. Image orientation is same as in Figure 1. The top row shows two images near rotational phase 0.71 for a periodicity of 17.1 hours (zero phase is chosen as 00:00 UT on September 30). The October 3 image indicates that its coma structure is less evolved than that of the October 1 image suggesting the periodicity should be >17.1 hours. In fact, the rotational phase of the October 3 image lags the October 1 image (taken at 05:19 UT) by a phase of 0.10 for a periodicity of 17.6 hours. The bottom row shows two images near rotational phase 0.07 for a periodicity of 17.1 hours. The October 4 image shows that it requires further evolution for a match. This image lags the corresponding October 1 image (taken at 11:22 UT) by a phase of 0.09 for a periodicity of 17.6 hours. Right: top and bottom rows show two pairs of images at identical rotational phases for a periodicity of 17.6 hours for the October run. Note the almost perfect repeatability of the coma morphology for these two image pairs. Zero phase was chosen as 00:00 UT on September 30. The images are ~39,000 km across. The PA and $\alpha$ are ~7° and ~45° respectively.

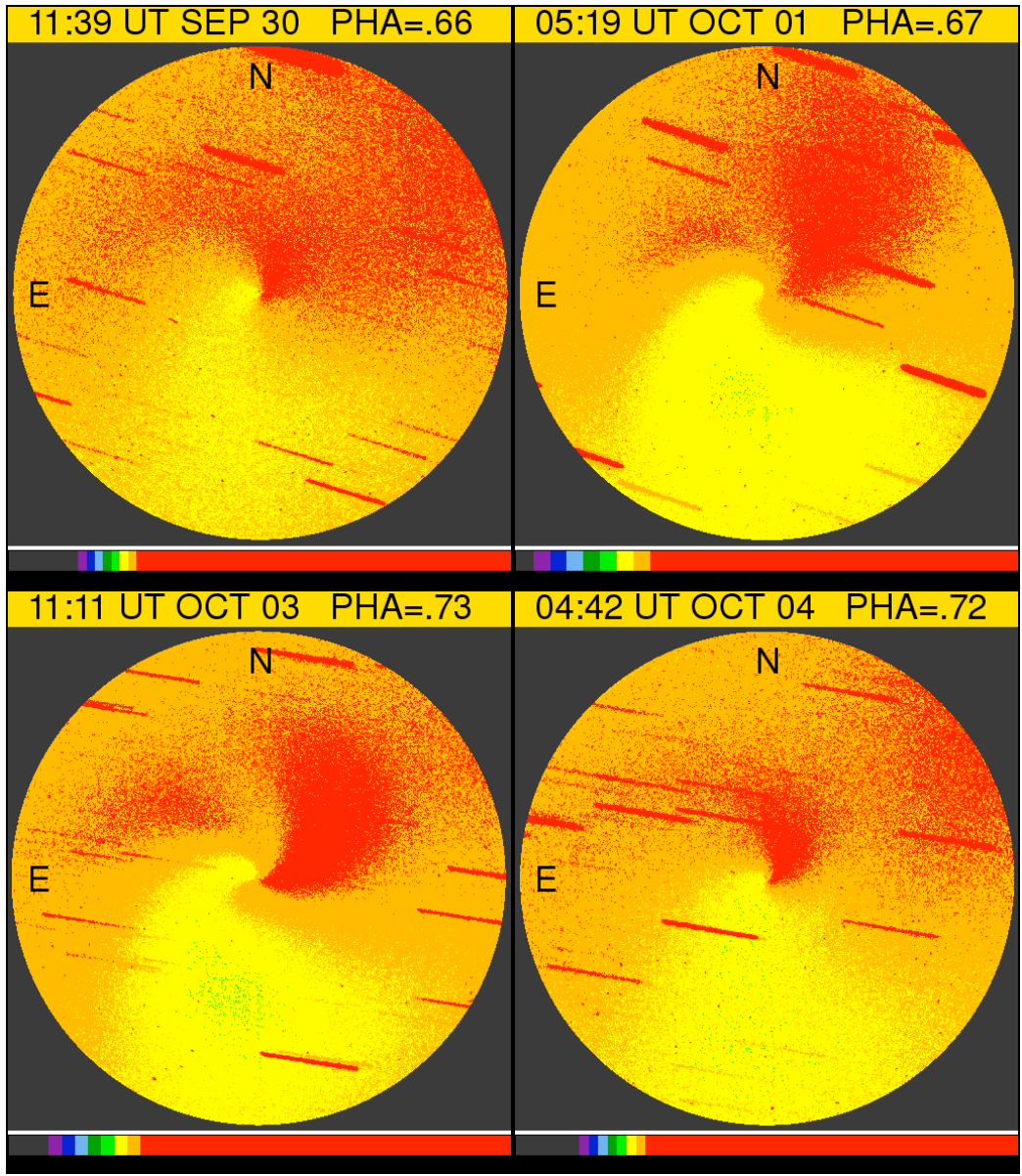

**Figure 4.** Two pairs of images showing the CN morphology for a rotational phase of ~0.66 (top row) and ~0.73 (bottom row) for a periodicity of 17.6 hours. The images in a pair are chosen from consecutive rotational cycles and despite the almost identical phase, there are slight differences in morphology. This behavior where most images of the same rotational phase from different rotational cycles show repeatability but a few images show minor differences is consistent with a low-excitation non-principal axis rotation. Corresponding changes in the sun and Earth directions for an image pair are ~1° and ~2° and therefore the geometry changes are negligible. The images are ~39,000 km across. Zero phase was chosen as 00:00 UT on September 30. The PA and $\alpha$ are ~7° and ~45° respectively.

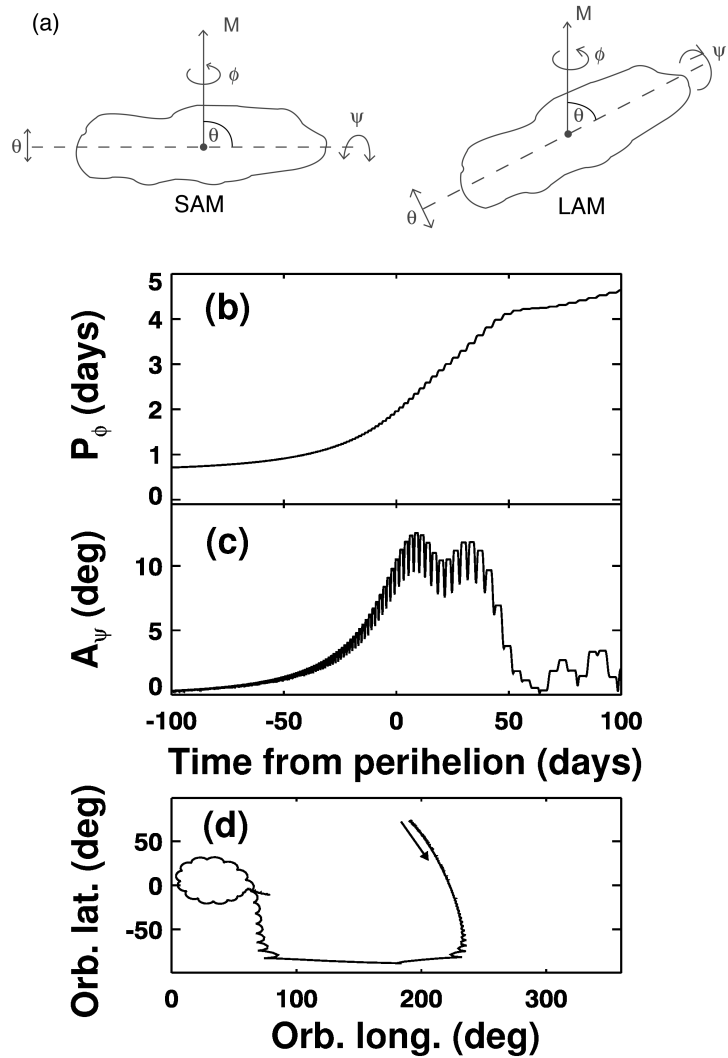

**Figure 5.** The two modes of non-principal axis rotational states (Julian 1987) are Short Axis Modes (SAMs) and Long Axis Modes (LAMs). For an elongated object like 103P/Hartley 2, from an observer's perspective, it makes sense to describe the component motions of rotation with respect to the long axis. Panel (a) shows the component motions for a SAM and a LAM. For a 103P/Hartley 2-like nucleus, the low-excitation rotational states have $\theta \approx 90°$ where $\theta$ is the angle between the RAMV, **M**, and the long axis. The variations in angle $\phi$ represent a circulatory motion of the long axis around the RAMV; variations in angle $\psi$ represent a librating roll (in the case of a SAM) or a circulatory roll (in the case of a LAM) of the long axis around itself; variations in angle $\theta$ represent a nutation of the long axis. The respective motions associated with variations in angles $\phi$, $\psi$, and $\theta$ are shown by arrows. The nutation amplitude is small enough to be ignored for a 103P/Hartley 2-like nucleus (see Appendix of Samarasinha and A'Hearn 1991). Panels (b-d) show the results from a numerical simulation to demonstrate that a 103P/Hartley 2-like nucleus could easily experience large changes to its rotational state over a timescale

of a few months. These plots are for demonstration purposes only and detailed calculations should be carried out once the shape model and the nuclear activity are determined from the EPOXI data. This particular simulation corresponds to a scenario where the dominant torque is due to an active region near the end of the long axis and the nucleus excites in to SAM non-principal axis states. Panel (b) shows the evolution of period, $P_\phi$ (i.e., the period for the circulatory motion in angle $\phi$), as a function of time. Panel (c) shows the amplitude of the librating roll, $A_\psi$, of the long axis for the excited SAM as a function of time. This simulation assumes a relaxed principal–axis state prior to the start of activity at large heliocentric distances. Panel (d) illustrates the temporal evolution of the RAMV in an orbit-based coordinate system of orbital latitude versus orbital longitude. The orbital longitude is measured from the sun-perihelion direction and advances along the orbital plane in the direction of the comet motion. In panels (b-d), the "wiggles" are due to "diurnal" variations.